\documentclass[conference]{IEEEtran}
\usepackage{amssymb,amsmath,amsfonts,amsthm,dsfont}
\usepackage[normalem]{ulem}
\usepackage{bbm}
\usepackage{bm}
\usepackage{physics}
\usepackage{graphicx,wasysym}
\usepackage{xcolor}
\usepackage{float}
\usepackage{mathtools}
\usepackage{subcaption}

\newcommand{\sgn}[1]{\text{sgn}\left\lbrack#1\right\rbrack}

\ifCLASSINFOpdf
\else
\fi

\begin{document}
%
\title{Robust quantum classifier with minimal overhead}

\author{\IEEEauthorblockN{Daniel K. Park\IEEEauthorrefmark{1}\textsuperscript{\textparagraph},
Carsten Blank\IEEEauthorrefmark{2}\textsuperscript{\textparagraph}, and
Francesco Petruccione\IEEEauthorrefmark{3}\IEEEauthorrefmark{4}}
\IEEEauthorblockA{\IEEEauthorrefmark{1}Sungkyunkwan University Advanced Institute of Nanotechnology,
Suwon, Korea\\ Email: dkp.quantum@gmail.com}
\IEEEauthorblockA{\IEEEauthorrefmark{2}Data Cybernetics, Landsberg, Germany\\
Email: blank@data-cybernetics.com}
\IEEEauthorblockA{\IEEEauthorrefmark{3}School of Chemistry and Physics, University of KwaZulu-Natal, Durban, South Africa}
\IEEEauthorblockA{\IEEEauthorrefmark{4}National Institute for Theoretical Physics (NITheP), KwaZulu-Natal, South Africa\\ Email: petruccione@ukzn.ac.za}}

\maketitle
\begingroup\renewcommand\thefootnote{}
\footnotetext{\textsuperscript{\textparagraph}Equal contribution}
\footnotetext{\copyright2021 IEEE}
\endgroup

\begin{abstract}
To witness quantum advantages in practical settings, substantial efforts are required not only at the hardware level but also on theoretical research to reduce the computational cost of a given protocol. Quantum computation has the potential to significantly enhance existing classical machine learning methods, and several quantum algorithms for binary classification based on the kernel method have been proposed. These algorithms rely on estimating an expectation value, which in turn requires an expensive quantum data encoding procedure to be repeated many times. In this work, we calculate explicitly the number of repetition necessary for acquiring a fixed success probability and show that the Hadamard-test and the swap-test circuits achieve the optimal variance in terms of the quantum circuit parameters. The variance, and hence the number of repetition, can be further reduced only via optimization over data-related parameters. We also show that the kernel-based binary classification can be performed with a single-qubit measurement regardless of the number and the dimension of the data. Finally, we show that for a number of relevant noise models the classification can be performed reliably without quantum error correction. Our findings are useful for designing quantum classification experiments under limited resources, which is the common challenge in the noisy intermediate-scale quantum era.
\end{abstract}


%
\IEEEpeerreviewmaketitle

\section{Introduction}
The theory of fault-tolerant quantum computing promises tremendous opportunities with clear quantum advantages for certain computational tasks~\cite{10.2307/2899535,zalka1998simulating,shor1999polynomial,PhysRevLett.103.150502_HHL_qBLAS}. However, the development of full-fledged quantum computing hardware remains a long-term prospect. On the road to building fault-tolerant quantum computers, noisy intermediate-scale quantum (NISQ) computers are expected to be available in the near future~\cite{Preskill2018quantumcomputingin,Google_QS,ion_trap_BM}. These quantum devices can execute only a limited size of quantum circuits reliably due to noise, but can surpass the capabilities of classical digital computers. An important issue in the NISQ era is to find problems and applications for which the NISQ technology can provide practical quantum advantage. Addressing this issue involves not only experimental efforts at the hardware level but also requires theoretical research to develop quantum algorithms while taking imperfections into account.

Quantum computing also has the potential to drastically improve machine learning tasks~\cite{wittek,QML-Biamonte,SupervisedQML,Dunjko_2018,QML_PRSA}. Quantum advantages in machine learning are expected naturally since quantum computers can reduce the computational cost exponentially for solving certain basic linear algebra problems~\cite{Nielsen:2011:QCQ:1972505,PhysRevLett.103.150502_HHL_qBLAS} that often appear as basic subroutines in machine learning tasks. Moreover, quantum computers can achieve exponential compression of data~\cite{PhysRevLett.100.160501,ffqram,9259210}. Full comprehension of which machine learning problems can be solved more efficiently with quantum algorithms remains as an important open problem.

A family of machine learning tasks for which quantum techniques are expected to outperform existing methods is the kernel-based classification~\cite{Havlicek2019,PhysRevLett.113.130503_qSQVM,PhysRevLett.122.040504,QML_Maria_Francesco,blank_quantum_2020,PARK2020126422}, which is a fundamental problem in pattern recognition. A common advantage of quantum computing utilized in these works is the ability to manipulate exponentially large quantum Hilbert space efficiently and evaluate the kernel function much faster than classical computers. The quantum machine learning algorithms presented in Refs.~\cite{QML_Maria_Francesco,blank_quantum_2020} are of particular interest in the NISQ era since they do not require expensive subroutines for solving the convex optimization problem of support vector machine. Furthermore, they are flexible in terms of the quantum data encoding method; either amplitude encoding~\cite{SupervisedQML} or quantum feature mapping~\cite{PhysRevLett.122.040504} can be used. In these algorithms, a classification score is evaluated by measuring an expectation value of certain observables, and hence repeating the same experiment multiple times is inevitable for a reliable statistics. However, due to the measurement postulate of quantum mechanics and the no-cloning theorem, a same input state must be created for every execution of the algorithm~\cite{Park_2019_forking}. Since computational cost for preparing an arbitrary quantum input state can be substantial depending on the structure of data to be encoded, reducing the number of repetition is essential~\cite{Park_2019_forking}, especially for NISQ computing.

In this work, we explicitly calculate the number of repetition necessary for estimating the classification score with a fixed precision. We discuss various generalizations that can be made to existing kernel-based quantum classifiers and calculate the variance of the estimator to show that the classifiers presented in Refs.~\cite{QML_Maria_Francesco,blank_quantum_2020} are indeed optimal with respect to the number of repetition. We also show that the kernel-based binary classification can be performed with a single-qubit measurement regardless of the number and the dimension of the data. This is particularly useful for the systems in which the measurement error is worse than the gate error. Furthermore, we show that the binary classifications can be performed reliably under certain noise models that are important in quantum information science without having to employ expensive quantum error correction; the classification will succeed by increasing the number of repetition quadratically with respect to a relevant error rate.


\section{Preliminaries}

\subsection{Binary classification}
Classification is a fundamental problem in machine learning. The goal of $L$-class classification is to infer the class label of an unseen data point $\tilde{x} \in \mathbbm{C}^N$, given a labelled data set $$\mathcal{D} = \left\{ (x_1, y_1), \ldots, (x_M, y_M) \right\} \subset \mathbbm{C}^N\times\{0,1,\ldots,L-1\}.$$ Although the data is real-valued in usual machine learning tasks, we allow complex-valued data without loss of generality by noting that various quantum data encoding schemes utilize the quantum Hilbert space. A famous example of encoding classical information as a quantum state is the amplitude encoding which represents a classical vector $\mathbf{x}_j = (x_{1j}, \ldots, x_{Nj})^T \in \mathbbm{C}^N$ as a quantum state in the following form,
\begin{equation}
\label{eq:amplitude_encoding_1}
    \ket{\mathbf{x}_j} := \frac{1}{\|\mathbf{x}_j\|} \sum_{i=1}^{N} x_{ij} \ket{i},
\end{equation}
using $\lceil \log_2(N)\rceil$ qubits. Similarly, a set of $M$ data points $\mathbf{x}_1, \ldots, \mathbf{x}_M$ can be encoded in $\lceil \log_2(NM)\rceil$ qubits as
\begin{equation}
\label{eq:amplitude_encoding_M}
    \frac{1}{\sqrt{\sum_{ij}|x_{ij}|^2}} \sum_{j=1}^{M}\sum_{i=1}^{N} x_{ij} \ket{i}\otimes\ket{j}.
\end{equation}

Hereinafter we focus on binary classification (i.e. $L=2$) like majority of works on quantum kernel-based classifiers since a multi-class classification can be constructed with binary classifiers by one versus all or one versus one scheme. In addition, we will omit the Kronecker product symbol ($\otimes$) whenever the meaning is clear (e.g. $\ket{i}\otimes\ket{j}= \ket{ij}$).

\subsection{Review of kernel-based quantum classifiers}

This work focuses on extending and improving the kernel-based quantum classifiers presented in Refs.~\cite{QML_Maria_Francesco,blank_quantum_2020}, since they are more suitable for NISQ computing as mentioned in the introduction. These algorithms are referred to as \textit{Hadamard-test classifier} (HTC) and \textit{swap-test classifier} (STC), respectively. 
Construction of these algorithms can be broken into two parts: preparation of a quantum state that encodes data in a specific form and expectation value measurement. These are explained in more detail below.

The Hadamard-test classifier encodes the dataset $\mathcal{D}$ in a quantum state as
		\begin{equation}
			\label{eq:htc_initial}
			\ket{\psi_h} = \frac{1}{\sqrt{2}} \sum_{j=1}^{M} \sqrt{a_j}\left(\ket{0}\ket{\mathbf{x}_j} + \ket{1}  \ket{\tilde{\mathbf{x}}} \right)\ket{y_j}\ket{j},
		\end{equation}
where $\ket{\mathbf{x}_j}$ and $\ket{\tilde{\mathbf{x}}}$ encodes classical training and test data vectors via an encoding of choice, and the label $y_j\in\lbrace 0,1 \rbrace$ is represented by the computational basis of the label qubit. Without loss of generality all inputs $\mathbf{x}_j$ and $\tilde{\mathbf{x}}$ are assumed to be normalized and have unit length. The subscript $h$ indicates that the state is for the Hadamard-test classifier. In Ref.~\cite{QML_Maria_Francesco}, the weights are uniform, i.e. $a_j = 1/M\;\forall j$, but it can be left as a variable to be optimized, similar to the treatment in support vector machines~\cite{PARK2020126422}. The measurement scheme utilizes a Hadamard-test, which applies a Hadamard gate on the ancilla qubit to interfere training and test data states. Finally, by measuring an expectation value of a two-qubit observable $\sigma_z^{(a)}\sigma_z^{(l)}$ on the ancilla qubit and the lable qubit, one obtains
\begin{equation}
	\label{eq:htc_score}
    \langle\psi_h|H^{(a)} \sigma_z^{(a)}\sigma_z^{(l)}H^{(a)}|\psi_h\rangle=\sum_{j=1}^M(-1)^{y_j}a_j\text{Re}\braket{\mathbf{x}_j}{\tilde{\mathbf{x}}},
\end{equation}
where the superscripts $a$ and $l$ indicate that the corresponding operator is acting on the ancilla qubit and the label qubit, respectively. From this equation, one can see that the kernel function in HTC is $k(\mathbf{x}_j,\tilde{\mathbf{x}})=\text{Re}\braket{\mathbf{x}_j}{\tilde{\mathbf{x}}}$, and Eq.~(\ref{eq:htc_score}) defines the classification score, which we denote by $f$, in an HTC. The HTC assigns a new label to the test data as $$\tilde{y} = \frac{1}{2}\left(1-\sgn{\sum_{j=1}^M(-1)^{y_j}a_j\text{Re}\braket{\mathbf{x}_j}{\tilde{\mathbf{x}}}}\right).$$

The HTC considers only a real part of the quantum state overlap. In order to fully exploit the ability of quantum computers to efficiently manipulate quantum states in the Hilbert space, it is desirable to construct a kernel that takes both real and imaginary parts of the quantum state into account. This motivated the birth of the swap-test classifier (STC). To see how the kernel function looks like in the STC, it is useful to express the initial state with the density matrix formalism. The initial state can be written as
\begin{equation}
\label{eq:stc_initial}
    \rho_s = \ketbra{0}{0}\otimes\sum_{j=1}^M \left( a_j\left(\tilde\rho\otimes\rho_j\right)^{\otimes k}\otimes\ketbra{y_j}{y_j}\otimes\ketbra{j}{j} \right),
\end{equation}
where the subscript $s$ indicates that the density matrix is for the swap-test classifier. The next step of the classifier is to apply the swap test
\begin{equation}
\label{eq:swap_test}
T_{s}=H^{(a)}\cdot\prod_{i=1}^k\texttt{swap}(t_i,d_i\vert a=1)\cdot H^{(a)},
\end{equation}
where $H^{(a)}$ represents a Hadamard gate applied to the ancilla qubit and $\texttt{swap}(t_i,d_i\vert a)$ represents a controlled-swap gate that exchanges an $i$th copy of test ($t_i$) and training ($d_i$) data if the ancilla qubit state is $a$. Finally, the expectation value measurement of a two-qubit observable $\sigma_z^{(a)}\sigma_z^{(l)}$ results in
\begin{equation}
\label{eq:stc_score}
    \langle \sigma_z^{(a)}\sigma_z^{(l)} \rangle =\sum_{j=1}^M(-1)^{y_j}a_j\Tr(\tilde\rho\rho_j)^k.
\end{equation}
Therefore, the kernel function in the STC is $k(\mathbf{x}_j,\tilde{\mathbf{x}})=\Tr(\tilde\rho\rho_j)^k$.
When the training and test data are given as a pure state, i.e. $\tilde\rho = \ketbra{\tilde{\mathbf{x}}}{\tilde{\mathbf{x}}}$ and $\rho_j = \ketbra{\mathbf{x}_j}{\mathbf{x}_j}$, then the kernel is reduced to $k(\mathbf{x}_j,\tilde{\mathbf{x}})=|\langle \tilde{\mathbf{x}}|\mathbf{x}_j\rangle|^{2k}$, which is the $k$th power of the quantum state fidelity. The classification score in an STC is given by Eq.~(\ref{eq:stc_score}). The STC assigns a new label to the test data as $$\tilde{y} = \frac{1}{2}\left(1-\sgn{\sum_{j=1}^M(-1)^{y_j}a_j\Tr(\tilde\rho\rho_j)^k}\right).$$

\section{Optimization}
We seek to minimize the computational resource overhead in the HTC and STC caused by the number of repetition necessary for estimating the classification score. The number of repetition can be calculated from the Chebyshev inequality 
\begin{equation}
\label{eq:Chebyshev}
    \Pr[|\mu - \langle \mathcal{M}\rangle | \ge \epsilon] \le \sigma^2/(k\epsilon^2) 
\end{equation}
where $\mu$ is an average value obtained from $k$ trials, $\langle \mathcal{M}\rangle$ is the expectation value to be estimated, and $\sigma^2=\Delta \mathcal{M} = \expval{\mathcal{M}^2} - \expval{\mathcal{M}}^2$ is the variance. Since the classifier only uses the sign of the expectation value, we can choose the precision to be 
\begin{equation}
    \epsilon = \langle\mathcal{M}\rangle/c
\end{equation}
for some constant $c>1$. Then the desired number of repetition goes as
\begin{equation}
\label{eq:num_shots}
    k=O(\sigma^2/\langle \mathcal{M}\rangle^2)
\end{equation}
to bound the error probability to a fixed constant.

In the following, we examine how to reduce $k$ by changing the circuit design of the classifier.

\subsection{General form of the classification score}
Suppose one measures an expectation value of an observable $\mathcal{M}_\lambda = \sigma_z^{(a)} \otimes A^{(l)}_\lambda$ for some arbitrary Hermitian operator $A_\lambda$. In addition, suppose the classical label $y_j$ is encoded in a logical label state $|\bar{y}_j\rangle$. Then it is straight-forward to see that the expectation value measured in either HTC or STC classification protocols becomes
\begin{equation}
\label{eq:expect_general}
    \langle\mathcal{M}_\lambda\rangle =\sum_{j=1}^M a_j \langle \bar{y}_j |A_\lambda|\bar{y}_j\rangle k(\mathbf{x}_j,\tilde{\mathbf{x}}),
\end{equation}
where the kernel function depends on whether the HTC or the STC is performed. From the above equation, one can see that the classification contrast can be increased by choosing $A$ and $|\bar{y}_j\rangle$ such that $\langle \bar{y}_j |A_\lambda|\bar{y}_j\rangle \in \lbrace -\lambda, \lambda\rbrace$ for some $\lambda>1$.

Now we choose an observable and the label qubit state based on the following rule:
\begin{equation}
    \sum_{i=1}^{\lambda} \sigma_z^{(i)} |l\rangle^{\otimes \lambda}=(-1)^{l}\lambda|l\rangle^{\otimes \lambda},\;l\in\lbrace 0,1\rbrace,
\end{equation}
where the superscript $i$ indicates that the Pauli operator is acting on the $i$th qubit. By setting $A_\lambda=\sum_{i=1}^{\lambda} \sigma_z^{(i)}$ and $|\bar{y}_j\rangle = |0\rangle^{\otimes\lambda}$ if $\mathbf{x}_j$ is labelled 0 and $|\bar{y}_j\rangle =|1\rangle^{\otimes\lambda}$ if $\mathbf{x}_j$ is labelled 1, one obtains 
\begin{equation}
\label{eq:expect_new}
    \langle\mathcal{M}_\lambda\rangle =\lambda\sum_{m=1}^M a_j (-1)^{y_j} k(\mathbf{x}_j,\tilde{\mathbf{x}}).
\end{equation}
Therefore, the expectation value to be estimated is scaled by a factor of $\lambda$ by increasing the number of label qubits by the same factor.
\subsection{Variance calculation}
\label{sec:variance}

We assume that we have an eigenstate decomposition (which we can find) of the observable $\mathcal{M}_\lambda$ such that
\begin{align}
    \mathcal{M}_\lambda = m_{0,\bar{0}} P_{0,\bar{0}} + m_{0,\bar{1}} P_{0,\bar{1}} + m_{1,\bar{0}} P_{1,\bar{0}} + m_{1,\bar{1}} P_{1,\bar{1}}
\end{align}
where $m_{i,\bar{j}} = (-1)^{i+\bar{j}} \lambda$ and projections $P_{i,\bar{j}} = \ketbra{i,\bar{j}}{i, \bar{j}}$ for $i,\bar{j}=0,1$. Gathering the statistics we compute
\begin{align}
    \expval{\mathcal{M}_\lambda} =&  m_{0,\bar{0}} p(0,\bar{0}) + m_{0,\bar{1}} p(0,\bar{1}) \nonumber \\
    & + m_{1,\bar{0}} p(1,\bar{0}) + m_{1,\bar{1}} p(1,\bar{1}) \nonumber \\
    =&  \lambda\left(p(0,\bar{0}) - p(0,\bar{1}) - p(1,\bar{0}) + p(1,\bar{1}) \right) \label{eq:expval_M}
\end{align}
as well as
\begin{align}
\label{eq:m_squared}
    \expval{\mathcal{M}^2_\lambda} =&  m^2_{0,\bar{0}} p(0,\bar{0}) + m^2_{0,\bar{1}} p(0,\bar{1})  \nonumber \\
    &+ m^2_{1,\bar{0}} p(1,\bar{0}) + m^2_{1,\bar{1}} p(1,\bar{1})  \nonumber \\ 
    =&  \lambda^2 \left(p(0,\bar{0}) + p(0,\bar{1}) + p(1,\bar{0}) + p(1,\bar{1}) \right)  \nonumber\\
    =& \lambda^2
\end{align}
with $p(i,\bar{j}) = \Tr(P_{i,\bar{j}} \ketbra{\Psi_f}{\Psi_f})$. 

By identifying Eq.~(\ref{eq:expval_M}) with Eq.~(\ref{eq:expect_new}) we see that the variance is given by
\begin{align}
\label{eq:variance}
    \sigma^2&=\expval{\mathcal{M}^2_\lambda} - \expval{\mathcal{M}_\lambda}^2 \nonumber \\
    &= \lambda^2\left(1 - \left(\sum_{j=1}^M a_j (-1)^{y_j} k(\mathbf{x}_j,\tilde{\mathbf{x}})\right)^2\right)
\end{align}
which means that at the boundary, i.e. when both classes are equally far away for a test datum, the variance is maximal.

Since both $\sigma^2$ and $\langle \mathcal{M}_\lambda\rangle^2$ goes as $\lambda^2$, the number of repetition is independent of $\lambda$ according to Eq.~(\ref{eq:num_shots}). In other words, the number of repetition cannot be reduced by increasing the number of label qubits. Therefore, the best strategy with respect to the measurement of label register is to use as small number of qubits as possible. This is simply done by using a single qubit to encode the label information (i.e. $\lambda=1$), and setting $A^{(l)}_\lambda = \sigma_z$.

\subsection{Skewness}
\label{sec:skewness}
The skewness is a measure how much the distribution is leaning towards one side from the mean. It is the third standardized moment and with respect to the mean $\expval{\mathcal{M}}$ and can be expressed as
\begin{align}
    \mu_3^c &= \frac{\expval{\mathcal{M}_\lambda^3} - 3\expval{\mathcal{M}_\lambda} \Delta \mathcal{M}_\lambda - \expval{\mathcal{M}_\lambda}^3}{\Delta \mathcal{M}_\lambda^{3/2}}.
\end{align}
For an arbitrary $\lambda$, the third moment is given by
\begin{align}
\label{eq:m_cubed}
    \expval{\mathcal{M}^3_\lambda} &=  m^3_{0,\bar{0}} p(0,\bar{0}) + m^3_{0,\bar{1}} p(0,\bar{1})  \nonumber \\
    &\qquad  + m^3_{1,\bar{0}} p(1,\bar{0}) + m^3_{1,\bar{1}} p(1,\bar{1})  \nonumber \\ 
    &=  \lambda^3 \left(p(0,\bar{0}) - p(0,\bar{1}) - p(1,\bar{0}) + p(1,\bar{1}) \right)  \nonumber\\
    &= \lambda^3 \left(\sum_{j=1}^M a_j (-1)^{y_j} k(\mathbf{x}_j,\tilde{\mathbf{x}})\right).
\end{align}
By combining the two equations above and using a notation $f =\sum_{j=1}^M a_j (-1)^{y_j} k(\mathbf{x}_j,\tilde{\mathbf{x}})$) we obtain
\begin{align}
    \mu_3^c &= \frac{\expval{\mathcal{M}_\lambda^3} - 3\expval{\mathcal{M}_\lambda} \Delta \mathcal{M} - \expval{\mathcal{M}_\lambda}^3}{\Delta \mathcal{M}_\lambda^{3/2}} \nonumber \\
    &= \frac{\lambda^3 f - 3\lambda f \lambda^2 (1-f^2) - \lambda^3 f^3}{\lambda^3 (1-f^2)^{3/2}} \nonumber \\
    &= -\frac{2f}{\sqrt{1-f^2}}.
\end{align}
Two remarks are to be made here. First, the skewness does not depend on $\lambda$. This result again favors the use of one-qubit register for encoding the label information. The second is that the skewness is negative with respect to the classification score $f$. This indicates that the probability density favors values whose absolute value is larger than $|\expval{\mathcal{M}_\lambda}|$. The skewness is zero only if $f=0$ with which the algorithm cannot classify the test data; the skewness will always be non-zero when the classifier can make a decision. The necessity of the asymmetry opens an interesting research direction towards designing quantum classifiers based on the mode (e.g. majority vote) instead of the mean.

\subsection{Generalization of the interfering circuit}
Further generalization can be made in both HTC and STC by using arbitrary single qubit rotation gates instead of the Hadamard gates for creating superposition in the beginning and interference at the end on the ancilla qubit. In this case, Eq.~(\ref{eq:htc_initial}) becomes
\begin{equation}
\label{eq:htc_initial_general}
\ket{\psi_h} = \sum_{j=1}^{M} \sqrt{a_j}\left(\cos\frac{\theta_0}{2}\ket{0}\ket{\mathbf{x}_j} + \sin\frac{\theta_0}{2}e^{i\phi}\ket{1}  \ket{\tilde{\mathbf{x}}} \right)\ket{y_j}\ket{j}.
\end{equation}
The Hadamard gate at the end of the circuit for interfering two subspaces spanned by the computational basis of the ancilla qubit is also replaced with an arbitary rotation around the $y$-axis of the Bloch sphere $R_y(\theta_1)=\cos(\theta_1/2)I-i\sin(\theta_1/2)\sigma_y$. This gate can be followed by an arbitrary rotation around the $z$-axis, but since we are also measuring in the $\sigma_z$-basis, we can neglect it as it does not alter the measurement result. For a depiction of this setup, confer to figure~\ref{fig:circuits}. Then the two-qubit expectation value measurement gives
\begin{align}
    \langle\sigma_{z}^{(a)}\sigma_z^{(l)}\rangle = & \sum_{j=1}^{M}a_j(-1)^{y_j}\big{(}\cos(\theta_0)\cos(\theta_1) -\sin(\theta_0)\sin(\theta_1)\nonumber\\
    & \times \left(\cos(\phi)\text{Re}\braket{\mathbf{x}_j}{\tilde{\mathbf{x}}}-\sin(\phi)\text{Im}\braket{\mathbf{x}_j}{\tilde{\mathbf{x}}}\right)\big{)}.
\end{align}
This equation shows that the imaginary part of the state overlap $\braket{\mathbf{x}_j}{\tilde{\mathbf{x}}}$ can also contribute to the classification result, unlike in the original HTC. We leave the use of this imaginary part for classification as an interesting future work, and only focus on the case where $\text{Im}\braket{\mathbf{x}_j}{\tilde{\mathbf{x}}}=0$ to mimic the original HTC.
\begin{figure*}[!t]
\centering
\subfloat[]{
\includegraphics[scale=0.5]{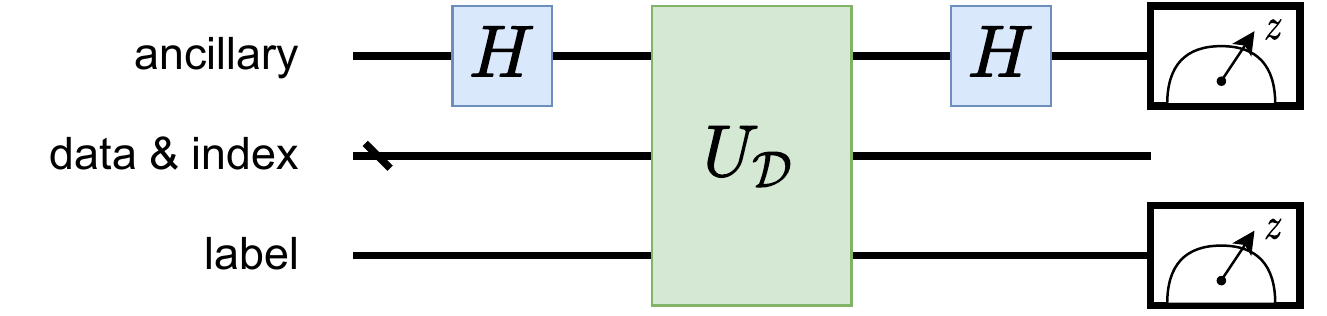}
}
\subfloat[]{
\includegraphics[scale=0.5]{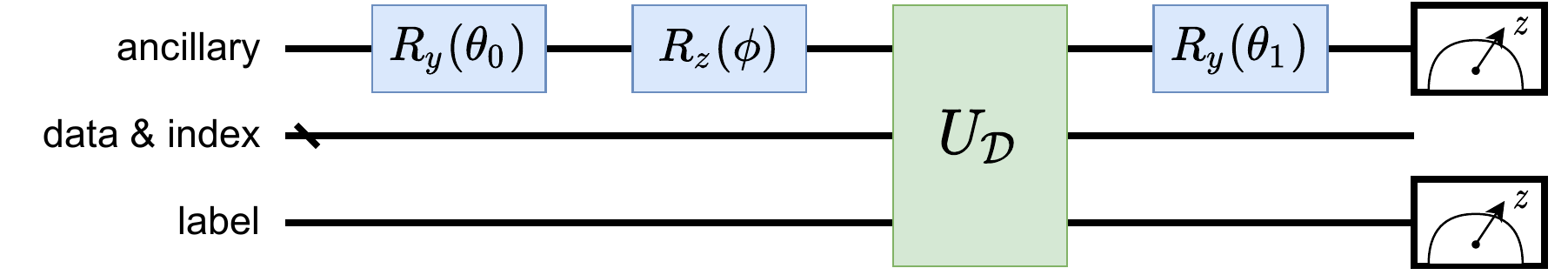}
}
\vspace{1cm}
\subfloat[]{
\includegraphics[scale=0.5]{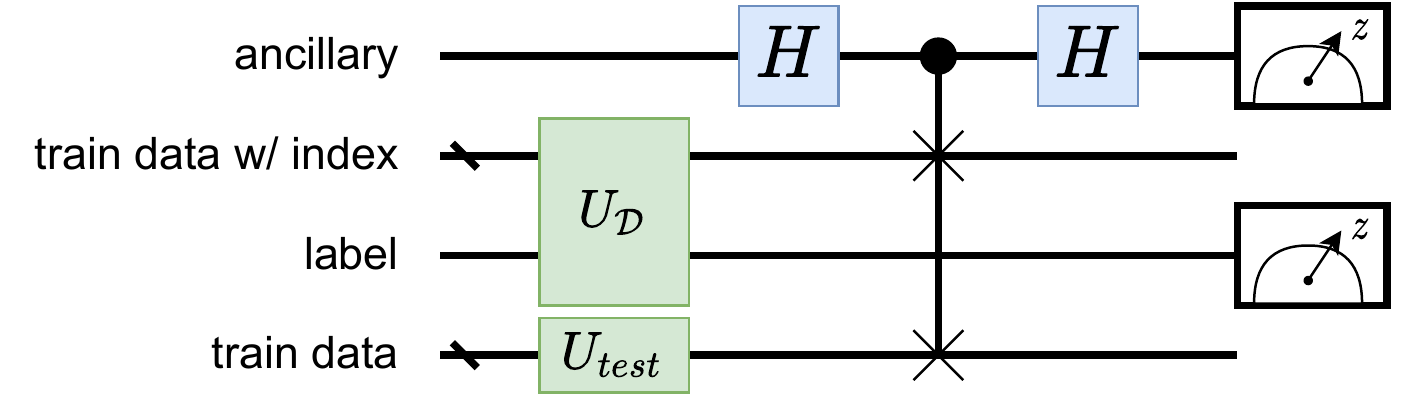}
}
\subfloat[]{
\includegraphics[scale=0.5]{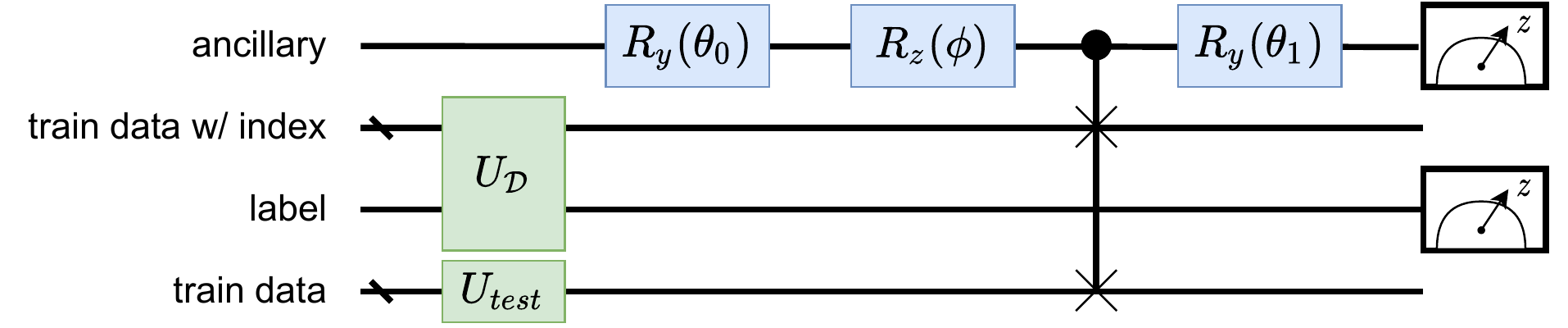}
}
\caption{\textbf{Classification Circuits.} (a) The Hadamard-test classifier depicted with the state preparation and the equal superpositionof the Hadamard. (b) This is altered to now have rotational gates with arbitrary angles instead of the two Hadamard gates. (c) and (d) are analogue but with the swap-test classifier.}
\label{fig:circuits}
\end{figure*}

Similarly, the general form of the expectation value for STC can be calculated. Without assuming $\text{Im}\braket{\mathbf{x}_j}{\tilde{\mathbf{x}}}=0$, it can be written as
\begin{align}
    \langle\sigma_{z}^{(a)}\sigma_z^{(l)}\rangle = & \sum_{j=1}^{M}a_j(-1)^{y_j}\big{(}\cos(\theta_0)\cos(\theta_1) -\sin(\theta_0)\sin(\theta_1)\nonumber\\
    & \times \cos(\phi)|\braket{\mathbf{x}_j}{\tilde{\mathbf{x}}}|^2\big{)}.
\end{align}

Now, it is straight-forward to see that $\langle (\sigma_z^{(a)}\sigma_z^{(l)})^2\rangle = \langle I \rangle =1 $. Thus the variance is simply $1-\langle (\sigma_z^{(a)}\sigma_z^{(l)})\rangle^2$, as expected from Eq.~(\ref{eq:variance}). Since we aim to minimize the variance, we want to maximize $f(\theta_0,\theta_1\phi) = \langle\sigma_z^{(a)}\sigma_z^{(l)}\rangle^2$; this is our objective function. For simplicity, we can consider a special cases where $\theta_0 = \pi/2$. In this case, the objective function becomes
$$
     \left\vert\sum_{j=1}^{M}a_j(-1)^{y_j}\sin(\theta_1)\cos(\phi)k(\mathbf{x}_j,\tilde{\mathbf{x}})\right\vert^2
$$
for both HTC and STC. This is maximized with respect to $\theta_1$ and $\phi$ if $|\sin(\theta_1)\cos(\phi)|$ = 1. One solution is $\sin(\theta_1) = 1$ and $\cos(\phi) = -1$, which is equivalent to setting the final gate on the ancilla qubit to be the Hadamard gate. 

More rigorous analysis can be done by calculating the first- and second-order partial derivatives of the objective function with respect to $\theta_0$, $\theta_1$, and $\phi$. After going through some laborious calculations, it can be shown that the critical point simultaneously satisfying
$$\frac{\partial f(\theta_0,\theta_1,\phi)}{\partial\theta_0} = 0,\; \frac{\partial f(\theta_0,\theta_1\phi)}{\partial\theta_1} = 0,\; \frac{\partial f(\theta_0,\theta_1\phi)}{\partial\phi} = 0$$
is a solution to $\cos(\theta_0^{*})=\cos(\theta_1^{*})=\sin(\phi^{*})=0$. This condition is satisfied in the original HTC and STC in which the creation of superposition and interference on the ancilla qubit is soley done by Hadamard gates. However, to be sure that the critical point really gives the local maximum of $f$, the second derivative test needs to be performed. To keep further discussions simple, we assume $\sin(\phi)=0$ so that the maximization is to be done only with respect to $\theta_0$ and $\theta_1$. In this case, the above critical point satisfies 
\begin{equation}
\label{eq:derivate_test1}
    \frac{\partial^2 f(\theta_0,\theta_1)}{\partial\theta_0^2} = \frac{\partial^2 f(\theta_0,\theta_1)}{\partial\theta_1^2} \le 0.
\end{equation}
The second derivatives are zero when $\sum_j{a_j}(-1)^{y_j}k(\mathbf{x}_j,\tilde{\mathbf{x}})=0$, in which case the classification is not possible anyways. Now, if 
\begin{equation}
\label{eq:derivate_test2}
\frac{\partial^2 f(\theta_0,\theta_1)}{\partial^2\theta_0}\frac{ \partial^2 f(\theta_0,\theta_1)}{\partial^2\theta_1} - \left(\frac{\partial^2 f(\theta_0,\theta_1)}{\partial\theta_0\partial\theta_1}\right)^2 > 0,    
\end{equation}
we can assure that the critical point above indeed yields the local maximum. This condition is satisfied if $\vert\sum_j{a_j}(-1)^{y_j}k(\mathbf{x}_j,\tilde{\mathbf{x}})\vert>\vert\sum_j{a_j}(-1)^{y_j}\vert$.

Having only two free parameters, another critical point is given when $\sin(\theta_0) = \sin(\theta_1) = 0$. This condition also satisfies Eq.~(\ref{eq:derivate_test1}), and hence a candidate for the maximizing solution. The condition in Eq.~(\ref{eq:derivate_test2}) is satisfied if $\vert\sum_j{a_j}(-1)^{y_j}k(\mathbf{x}_j,\tilde{\mathbf{x}})\vert<\vert\sum_j{a_j}(-1)^{y_j}\vert$. However, if $\sin(\theta_0) = \sin(\theta_1) = 0$ is used, then the classification outcome becomes $\langle\sigma_z^{(a)}\sigma_z^{(l)}\rangle = \sum_j{a_j}(-1)^{y_j}$, and hence does not construct the classification based on given dataset. Thus we conclude that this is not a suitable solution. Therefore, we argue that the best suited solution is given by the condition $\cos(\theta_0^{*})=\cos(\theta_1^{*})=\sin(\phi^{*})=0$, which is satisfied with the use of Hadamard gates. The variance is given as Eq.~(\ref{eq:variance}), and this is the optimal value if $\vert\sum_j{a_j}(-1)^{y_j}k(\mathbf{x}_j,\tilde{\mathbf{x}})\vert>\vert\sum_j{a_j}(-1)^{y_j}\vert$.

\section{Robustness to noise}
Imperfections are unavoidable in the implementation of quantum algorithms. The theory of quantum error correction and fault-tolerance guarantees that the quantum computation can be performed reliably under noise at the cost of increasing the quantum resources (e.g. qubits and gates) as long as the physical error rate is below certain threshold value. The resource overhead is larger if the physical error rate is larger, even if it is below the fault-tolerance threshold. Typical NISQ devices will not have enough number of qubits to perform an arbitrary fault-tolerant quantum computation. Therefore, minimizing the resource overhead for fighting against noise is of critical importance. In the following, we show that under certain relevant noise models, the quantum binary classification can be performed reliably without quantum error correction while increasing the number of repetition only quadratically with respect to an effective error rate.

\subsection{Classifier with a single-qubit measurement}
\begin{figure}[!t]
\centering
\subfloat[\label{fig:2a}]{
\includegraphics[scale=0.5]{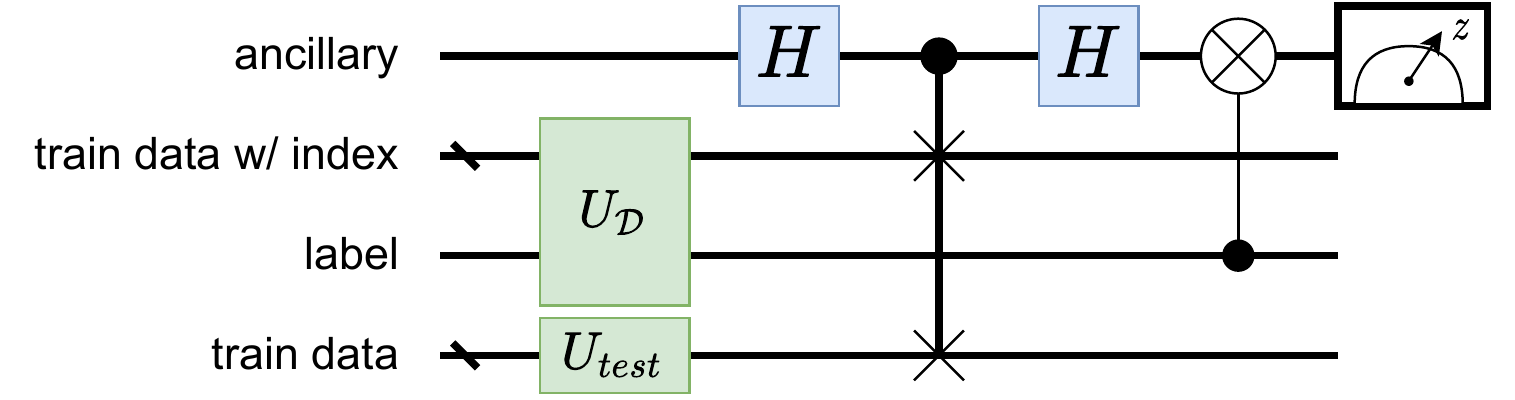}
}
\vspace{0.1cm}
\subfloat[\label{fig:2b}]{
\includegraphics[scale=0.62]{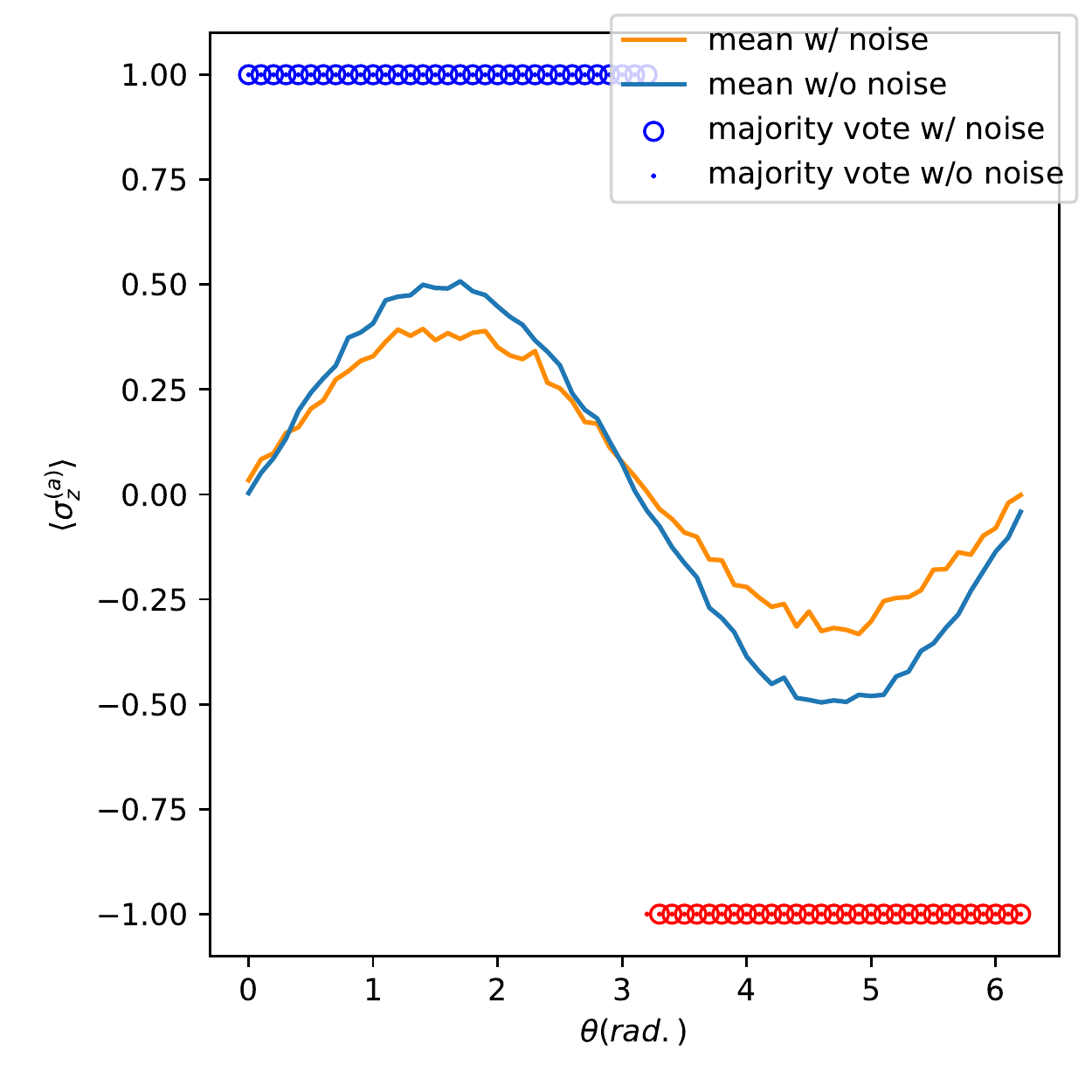}
}
\caption{\textbf{One Qubit Measurement.} (a) By adding a CNOT gate with the label qubit as control and the ancillary qubit as target, one can apply a one qubit z-measurement only. (b) Simulations of the one-qubit-measurement scheme simulated with and without noise. We compare the mean and the majority voting as the classification rule, and show that the majority voting is more favorable in practice. These results are obtained from 8192 shots and the noisy simulations are with respect to an IBM quantum devices called \texttt{ibmq\_rome}.}
\label{fig:circuits}
\end{figure}

Both HTC and STC rely on measuring an expectation value of a two-qubit observable $\sigma_z^{(a)}\sigma_z^{(l)}$. The same outcome can be obtained by measuring an expectation value of a single-qubit observable after an additional two-qubit gate. We introduce a notation for the controlled bit-flip (CNOT) operation as $I^{(i)}\otimes \ketbra{0}{0}^{(j)} + \sigma_x^{(i)}\otimes \ketbra{1}{1}^{(j)} = cX(i|j)$ to mean that the bit-flip operation is applied to a target qubit $i$ if the state of the control qubit $j$ is $|1\rangle$. Then the following property holds:
\begin{equation}
    cX(i|j)^{\dagger}\left(\sigma_z^{(i)}\otimes I^{(j)}\right)cX(i|j) \nonumber  = \sigma_z^{(i)}\otimes \sigma_z^{(j)}.
\end{equation}
With the above it is trivial to see that
\begin{align}
    \langle \sigma_z^{(a)}\sigma_z^{(l)}\rangle & = \Tr( \rho_f \sigma_z^{(a)}\sigma^{(l)}) \nonumber \\
    & = \Tr( \rho_f cX(a|l)^\dagger\sigma_z^{(a)}cX(a|l)) \nonumber \\
    & = \Tr( cX(a|l) \rho_f cX(a|l)^\dagger\sigma_z^{(a)}) \nonumber \\
    & = \Tr( \tilde{\rho}_f \sigma_z^{(a)}),
\end{align}
where $\rho_f$ is the density matrix representation of the final state of either the HTC or the STC quantum circuit, and $\tilde{\rho}_f$ is the modified final state in these circuits obtained by applying the CNOT operation on the ancilla qubit with the label qubit as the control. Therefore, the HTC and STC algorithms can be performed with a single-qubit measurement with addition of a CNOT gate. The quantum circuit with this modification is depicted in Fig.~\ref{fig:2a}

As a simple demonstration, we took the toy example from Ref.~\cite{blank_quantum_2020} and simulated the quantum binary classification protocol with the one-qubit measurement scheme. The example data set consists of two training data and one test data as
	\begin{align}
		\label{eqn:data_example}
		&\ket{\mathbf{x}_1} = \frac{i}{\sqrt{2}} \ket{0} + \frac{1}{\sqrt{2}} \ket{1},\; y_1 = 0,\nonumber\\
		&\ket{\mathbf{x}_2} = \frac{i}{\sqrt{2}} \ket{0} - \frac{1}{\sqrt{2}} \ket{1},\; y_2 = 1,\nonumber \\
		&\ket{\tilde{\mathbf{x}}} = \cos{\frac{\theta}{2}} \ket{0} - i \sin{\frac{\theta}{2}} \ket{1}.
	\end{align}
Simulations are carried out in two sets. First, we assume an ideal implementation without noise to verify the idea. Then we performed simulations with a realistic noise model. The noisy simulations implemented quantum circuits with gate decomposition (i.e. transpilation) given by a five-qubit IBM quantum device available called \texttt{ibmq\_rome} and its noise model. We also tested the use of majority vote for classification in place of the expectation value as suggested in Sec.~\ref{sec:skewness}. Each quantum circuit is repeated 8192 times to gather the measurement statistics. The simulation results are shown in Fig.~\ref{fig:2b}.

\subsection{Effect of noise}

As a simple noise model, let us consider the depolarizing error that acts only on the ancilla qubit at the end of the quantum circuit before the measurement. The effect of the single-qubit depolarizing channel acting on the final state $\tilde\rho_f$ can be studied with the Kraus representation as
\begin{equation}
    \mathcal{E}_d(\tilde\rho_f) = \sum_{k=1}^{4} E_k^{(a)} \tilde\rho_f \left(E_k^{(a)}\right)^{\dagger},
\end{equation}
where the set of Kraus operators are given as
$$ E = \left\lbrace \sqrt{1-\frac{3p}{4}} I , \sqrt{\frac{p}{4}} \sigma_x, \sqrt{\frac{p}{4}} \sigma_y, \sqrt{\frac{p}{4}} \sigma_z \right\rbrace$$ with a depolarizing error rate $0\le p \le 1$.
Then one can calculate the expectation value under this noise model as
\begin{align}
    \Tr(\mathcal{E}_d(\tilde\rho_f)\sigma_z^{(a)}) & = \Tr(\sum_{k=1}^{4} E_k^{(a)} \tilde\rho_f \left(E_k^{(a)}\right)^{\dagger} \sigma_z^{(a)})\nonumber \\
    & =\sum_{k=1}^{4} \Tr( \tilde\rho_f E_k^{(a)} \sigma_z^{(a)} E_k^{(a)})\nonumber \\
    & = (1 - p)  \Tr( \tilde\rho_f \sigma_z^{(a)}),
\end{align}
where the last line is obtained by using the commutation relations of the Pauli operators. This equation shows that under the single-qubit depolarizing noise model, the expectation value to be measured in HTC and STC are reduced by a factor of $1-p$, but importantly, it will not change the sign of the expectation value. This has an imperative consequence; the single-qubit depolarizing noise acting on the ancilla qubit of the final state can be easily mitigated since the classification only uses the sign of the measurement outcome. The same level of the classification accuracy as that of the noiseless case can be achieved by repeating the measurement $O(1/(1-p)^2)$ times. This result is deduced again from the Chebyshev inequality in Eq.~(\ref{eq:Chebyshev}).

Now we extend the error model to be an arbitrary Pauli channel acting on all qubits, which is the most basic noise channel ubiquitous in quantum information science~\cite{RevModPhys.87.307}. That is,
\begin{equation}
    \mathcal{E}_{p} = \sum_{j = 1}^{4^{n}} c_j P_j \tilde\rho_f P_j^{\dagger},
\end{equation}
where $P_j\in\lbrace I,\sigma_x,\sigma_y,\sigma_z\rbrace^{\otimes n}$ is an element in the set of $n$-qubit Pauli operators, and $c_j\in\mathbbm{R}$, $c\ge 0$, and $\sum_{j=1}^{4^n}c_j = 1$. Note that $P_j$ can be written as $P_j = P_j^{(1)}\otimes P_j^{(2)}\otimes \ldots \otimes P_j^{(n)}$. Under this noise model, the expectation value of interest becomes
\begin{align}
    & \Tr(\mathcal{E}_p(\tilde\rho_f)\sigma_z^{(a)})  =\Tr( \sum_{j}^{4^n} c_j P_j\tilde\rho_f P_j\sigma_z^{(a)})\nonumber \\
    & =\sum_{j}^{4^n} c_j \Tr( \tilde\rho_f P_j^{(a)}\sigma_z^{(a)} P_j^{(a)}\otimes P_j^{(2)}P_j^{(2)}\otimes\ldots\otimes P_j^{(n)}P_j^{(n)})\nonumber \\
    & = \sum_{j}^{4^n} c_j \Tr( \tilde\rho_f P_j^{(a)}\sigma_z^{(a)} P_j^{(a)})\nonumber \\
    & = (C_{I}+C_{\sigma_z}-C_{\sigma_x}-C_{\sigma_y})\Tr(\tilde\rho_f\sigma_z^{(a)}),
\end{align}
where $C_i = \sum_{j | P_{j}^{(a)} = i} c_j$. Therefore, an arbitrary Pauli noise channel acting on the final state of the classifier only scales the expectation value by a constant factor determined by the noisy process. As long as this factor is positive, the classifier can be made robust to noise by repeating the measurement $O(1/(C_{I}+C_{\sigma_z}-C_{\sigma_x}-C_{\sigma_y})^2)$ times.

\section{Conclusion}
Kernel-based quantum classification algorithms are promising candidates for NISQ applications. These algorithms are based on measuring an expectation value of an observable, which requires an experiment to be repeated many times to provide a good estimate. In this work, we investigated the possibilities and strategies to reduce this extra resource overhead. We showed explicitly that the variance in Hadamard-test and the swap-test circuits are optimal in terms of the quantum circuit design. In addition to the variance analysis, we calculated skewness and argued that the mode can be a statistically better quantity to measure for classification. We also explicitly calculated the number of repetition necessary to succeed a quantum classification protocol under relevant noise models and showed that it can be performed correctly without quantum error correction. The findings presented in this work are useful for designing NISQ experiments to perform quantum classification protocols with minimal overhead and robustness to noise.

\section*{Acknowledgment}
This research is supported by the National Research Foundation of Korea (Grant No. 2019R1I1A1A01050161), Quantum Computing Development Program (Grant No. 2019M3E4A1080227), and the South African Research Chair Initiative of the Department of Science and Technology and the National Research Foundation.

\bibliographystyle{IEEEtran}
\bibliography{references}



\end{document}